\documentclass{article} % For LaTeX2e
\usepackage{nips11submit_e,times,subfigure,ctable,amsmath,amsfonts,amssymb,cite,graphicx,caption,mathabx,amsthm,enumerate}

\newcommand{\X}{{\mathcal{X}}}
\newcommand{\Y}{{\mathcal{Y}}}

\newcommand{\data}{{\mathcal D}}

\newcommand{\cL}{{\mathfrak L}}
\newcommand{\bR}{{\mathbb R}}
\newcommand{\bI}{{\mathbb I}}
\newcommand{\cF}{{\mathcal F}}
\newcommand{\cV}{{\mathcal V}}

\newcommand{\fm}{{\mathfrak m}}

\newcommand{\bE}{{\mathbb E}}

\newcommand{\radem}{{\mathcal R}}

\title{Information, learning and falsification}

\author{
David Balduzzi\\
Max Planck Institute for Intelligent Systems, T{\"u}bingen, Germany. \\
\texttt{david.balduzzi@tuebingen.mpg.de}
}

\nipsfinalcopy % Uncomment for camera-ready version

\newtheorem{thm}{Theorem}

\newtheorem{propn}[thm]{Proposition}

\begin{document}
\maketitle

%%%%%%%%%%%%%%%%%%%%%%%%%%%%%%%%%%%%%%%%%%%%%%%%%%%%%%%%%%%%%%%%%%%%%%%%
%\section*{Extended Abstract}

There are (at least) three approaches to quantifying information. The first, algorithmic information or Kolmogorov complexity, takes events as  strings and, given a universal Turing machine, quantifies the information content of a string as the length of the shortest program producing it \cite{li:2008}.  The second, Shannon information, takes events as belonging to ensembles and quantifies the information resulting from observing the given event in terms of the number of alternate events that have been ruled out \cite{shannon:48}. The third, statistical learning theory, has introduced measures of capacity that control (in part) the expected risk of classifiers \cite{vapnik:98}. These capacities quantify the expectations regarding future data that learning algorithms embed into classifiers.

Solomonoff and Hutter have applied algorithmic information to prove remarkable results on universal induction. Shannon information provides the mathematical foundation for communication and coding theory. However, both approaches have shortcomings. Algorithmic information is not computable, severely limiting its practical usefulness. Shannon information refers to ensembles rather than actual events: it makes no sense to compute the Shannon information of a single string -- or rather, there are many answers to this question depending on how a related ensemble is constructed. Although there are asymptotic results linking algorithmic and Shannon information, it is unsatisfying that there is such a large gap -- a difference in kind -- between the two measures. 

This note describes a new method of quantifying information, effective information, that links algorithmic information to Shannon information, and also links both to capacities arising in statistical learning theory \cite{bt:08, balduzzi:11b}. After introducing the measure, we show that it provides a non-universal analog of Kolmogorov complexity. We then apply it to derive basic capacities in statistical learning theory: empirical VC-entropy and empirical Rademacher complexity. A nice byproduct of our approach is an interpretation of the explanatory power of a learning algorithm in terms of the number of hypotheses it falsifies \cite{popper:59}, counted in two different ways for the two capacities. We also discuss how effective information relates to information gain, Shannon and mutual information.

%%%%%%%%%%%%%%%%%%%%%%%%%%%%%%%%%%%%%%%%%%%%%%%%%%%%%%%%%%%%%%%%%%%%%%%%
\subsubsection*{Effective information}
Any physical system, at any spatiotemporal scale, is an input/output device. For simplicity, we only model memoryless systems with finite input $\X$ and output $\Y$ alphabets. The probability that system $\fm$ outputs $y\in \Y$ given input $x\in\X$ is encoded in Markov matrix $p_\fm(y|x)$. 

The effective information generated when system $\fm$ outputs $y$ is computed as follows. First, let the \emph{potential repertoire} $p_{unif}(X)$ be the input set equipped with the uniform distribution. Next, compute the \emph{actual repertoire} via Bayes' rule
\begin{equation}
	\label{e:actual}
	\hat{p}(X|y):= p\big(y|do(x)\big)\cdot\frac{p_{unif}(x)}{p_\fm(y)},
\end{equation}
where $p_\fm(y)=\sum_x p_\fm\big(y|do(x)\big)\cdot p_{unif}(x)$ and $do(\cdot)$ refers to Pearl's interventional calculus \cite{pearl:00}. \emph{Effective information} is the Kullback-Leibler divergence between the two repertoires
\begin{equation}
	ei(\fm,y) :=  D\big[\hat{p}_\fm(X|y)\,\big\|\,p_{unif}(X)\big].
	\label{e:ei}
\end{equation}
For a deterministic function $f:\X\rightarrow\Y$, the actual repertoire and effective information are
\begin{equation}
	\label{e:deterministic}
	\hat{p}_f(x|y)=\left\{\begin{matrix}
		\frac{1}{|f^{-1}(y)|} & \text{if }f(x)=y\\
		0 & \text{else}
	\end{matrix}\right.
	\,\,\,\,\,\,\text{ and }\,\,\,\,\,
	ei(f,y) = \log_2|\X|-\log_2|f^{-1}(y)|.
\end{equation}
The support of the actual repertoire is the pre-image $f^{-1}(y)$. Elements in the pre-image all have the same probability since they cannot be distinguished by the function $f$. Effective information quantifies the size of the pre-image relative to the input set -- the smaller (``sharper'') the pre-image, the higher $ei$.
 
%%%%%%%%%%%%%%%%%%%%%%%%%%%%%%%%%%%%%%%%%%%%%%%%%%%%%%%%%%%%%%%%%%%%%%%%
\subsubsection*{Algorithmic information}
We show that effective information is a non-universal analog of Kolmogorov complexity. Given universal Turing machine $T$, the (unnormalized) \emph{Solomonoff prior} probability of string $s$ is
\begin{equation}
	\label{e:solomonoff}
	p_T(s) := \sum_{\{i|T(i)= s\bullet\}} 2^{-\text{len}(i)},
\end{equation}
where the sum is over strings $i$ that cause $T$ to output $s$ as a prefix, where no proper prefix of $i$ outputs $s$, and $\text{len}(i)$ is the length of $i$. \emph{Kolmogorov complexity} is $K(s):=-\log_2 p_T(s)$. Kolmogorov complexity is usually defined as the shortest program on a universal prefix machine that produces $s$. The two definitions coincide up to additive constant by Levin's Coding Theorem \cite{li:2008}.

Replace universal Turing machine $T$ with deterministic system $f:\X\rightarrow \Y$. All inputs have $\text{len}(x)=\log_2|\X|$ in the optimal code for the uniform distribution on $\X$. Define the \emph{effective probability} of $y$ as
\begin{equation}
	\label{e:rad_dist}
	p_f(y) =
	\sum_{\left\{x|f(x)=y\right\}}2^{-\text{len}(x)}=
	\left\{\begin{matrix}
		\frac{|f^{-1}(y)|}{|\X|} & 
		\mbox{ if }y\in f(\X) 
		\vspace{1mm} \\ 
		0 & \mbox{ else.}
	\end{matrix}\right.
\end{equation}
Note that $p_f(y)$ is a special case of $p_\fm(y)$, as defined after Eq.~\eqref{e:actual}. The effective distribution is thus a non-universal analog of the Solomonoff prior, since it is computed by replacing universal Turing machine $T$ in Eq.~\eqref{e:solomonoff} with deterministic physical system $f:\X\rightarrow \Y$.

In the deterministic case, effective information turns out to be $ei(f,y)=-\log_2 p_f(y)$, analogously to Kolmogorov complexity. Effective information is non-universal -- but computable -- since it depends on the choice of $f$.

%%%%%%%%%%%%%%%%%%%%%%%%%%%%%%%%%%%%%%%%%%%%%%%%%%%%%%%%%%%%%%%%%%%%%%%%
\subsubsection*{Statistical learning theory}

This section uses a particular deterministic function, learning algorithm $\cL_{\cF,\data}$, to connect effective information and the effective distribution to statistical learning theory.

Given finite set $\X$, let \emph{hypothesis space} $\Sigma_\X=\big\{\sigma:\X\rightarrow\pm1\big\}$ contain all labelings of elements of $\X$. Now, given a set of functions $\cF\subset\Sigma_\X$ and unlabeled data $\data\in\X^l$, define \emph{learning algorithm} (empirical risk minimizer)
\vspace{-2mm}
\begin{equation}
	\label{e:learn}
	\cL_{\cF,\data}:\Sigma_\X\longrightarrow \bR:\hat{\sigma}\mapsto \epsilon=\min_{f\in \cF} \frac{1}{l}\sum_{k=1}^l \bI\left[f(d_k)\neq\hat{\sigma}(d_k)\right].
\end{equation}
The learning algorithm takes a labeling of the data as input and outputs the empirical risk of the function that best fits the data. We drop subscripts from the notation $\cL$ below. 

Define empirical VC-entropy in \cite{vapnik:98}) as $\cV(\cF,\data):=\log_2 |q_\data(\cF)|$ where $q_\data:\cF\rightarrow \bR^l:f\mapsto \big(f(d_1)\ldots f(d_l)\big)$. Also define empirical Rademacher complexity as
\begin{equation*}
	\radem(\cF,\data) = \frac{1}{|\Sigma|}
	\sum_{\sigma\in\Sigma}\left[\sup_{f\in\cF}\frac{1}{l}\sum_{k=1}^l\sigma(d_k)\cdot f(d_k)\right].
\end{equation*}
These capacities can be used to bound the expected risk of classifiers, see \cite{boucheron:00, bousquet:04} for details. The following propositions are proved in \cite{balduzzi:11b}:
\begin{propn}[effective information ``is'' empirical VC-entropy]
	\begin{equation*}
		ei(\cL,0) = - \log_2 p_{\cL}(0) = l - \cV(\cF,\data)
	\end{equation*}
\end{propn}

\begin{propn}[expectation over $p_\cL(\epsilon)$ ``is'' empirical Rademacher complexity]
	\begin{equation*}
		\bE[\epsilon\,|\,p_{\cL}] = \sum_{\epsilon\in\bR} \epsilon\cdot p_\cL(\epsilon)=\frac{1}{2}\big(1- \radem(\cF,\data)\big)
	\end{equation*}
\end{propn}
Thus, replacing the universal Turing machine with learning algorithm $\cL_{\cF,\data}$ we obtain that our analog of Kolmogorov complexity, the effective information of output $\epsilon=0$, is essentially empirical VC-entropy. Moreover, the expectation of the analog of the Solomonoff distribution is essentially Rademacher complexity. 

The two quantities $ei(\cL,0)$ and $\bE[\epsilon\,|\,p_{\cL}]$ are measures of explanatory power: as they increase, so expected future performance improves. By Eq.~\eqref{e:deterministic}, the effective information generated by $\cL$ is
\begin{equation}
	\label{e:falsify1}
	ei(\cL,0) = \underbrace{\log_2\left|\Sigma\right|}_{\mbox{total \# hypotheses}} 
	- \underbrace{\log_2|\cL^{-1}(0)|}_{\mbox{\# hypotheses }\cL\mbox{ fits}}		
 	= \Big(\mbox{\# hypotheses }\cL\mbox{ falsifies}\Big),
\end{equation}
where hypotheses are counted after logarithming. Effective information, which relate to VC-entropy, counts the number of hypotheses  the learning algorithm falsifies when it fits labels perfectly, without taking into account how often they are wrong. Similarly, see \cite{balduzzi:11b} for details, the expectation is
\begin{equation}
	\label{e:falsify2}
	\begin{matrix}
	\sum_\epsilon p_\cL(\epsilon)\cdot \epsilon 
	= \sum_\epsilon \Big(\mbox{fraction of hypotheses }\cL\mbox{ falsifies}\Big)\cdot \Big(\mbox{on fraction }\epsilon\mbox{ of the data}\Big).
	\end{matrix}
\end{equation}
Expected $\epsilon$, which relates to Rademacher complexity, looks at the average behavior of the learning algorithm, averaging over the fractions of hypotheses falsified, weighted by how much of the data they are falsified on.

The bounds proved in \cite{vapnik:98, boucheron:00, bousquet:04}, which control the expected future performance of the classifier minimizing empirical risk, can therefore be rephrased in terms of the number of hypotheses falsified by the learning algorithm, Eqs~\eqref{e:falsify1} and \eqref{e:falsify2}, suggesting a possible route towards rigorously grounding the role of falsification in science \cite{popper:59}.

%%%%%%%%%%%%%%%%%%%%%%%%%%%%%%%%%%%%%%%%%%%%%%%%%%%%%%%%%%%%%%%%%%%%%%%%
\subsubsection*{Shannon information}

We relate effective information to Shannon and mutual information. 

Suppose we have model $\fm$ that generates data $d\in D$ with probability $p_\fm(d|h)$ given hypothesis $h\in H$. For prior distribution $p(H)$ on hypotheses, the information gained by observing $d$ is
\begin{equation}
	\label{e:inf_gain}
	D\big[p_\fm(H|d)\,\big\|\, p(H)\big].
\end{equation}
Kullback-Leibler divergence $D[p\|q]$ can be interpreted as the number of Y/N questions required to get from $q$ to $p$. Thus, Eq.~\eqref{e:inf_gain} quantifies how many Y/N questions the model answers about the hypotheses using the data.

Effective information, Eq.~\eqref{e:ei}, quantifies the information gained when physical system $\fm$ outputs $y$. Rather than inferring on hypotheses, the system, by producing an output, specifies probabilistic constraints on what its input must have been. Effective information uses the uniform (maximum entropy) prior since any other prior would insert additional data not belonging to the system -- the prior is something else, \emph{on top of} $\fm$. However, this restriction is not essential and will be dropped for the remainder of this section.

Consider the following scenario. We have $X$ and $X'$ are isomorphic, and a deterministic physical system $c:X\rightarrow X'$ that \emph{copies} its inputs, mapping $x_k\mapsto x_k'$ for example. Given prior $p(X)$, the effective information generated is
\begin{equation*}
	ei\big(p(X),c,x_k'\big):=D\big[p_c(X|x_k')\,\big\|\,p(X)\big] = D\big[\delta_{x_k}\,\big\|\,p(X)\big] = -\log_2 p(x_k),
\end{equation*}
the surprise of $x_k$. It follows that Shannon information is expected effective information
\begin{equation*}
	H(X) = \bE\Big[ei\big(p(X),c,x_k'\big)\,\Big|\, p_c(X')\Big].
\end{equation*}
More generally, if we are given noisy memoryless channel $\fm$ from $\X$ to $\Y$ with distribution $p(X)$ on $X$, then mutual information is the expectation
\begin{equation*}
	I(X;Y) = \bE\Big[ei(p(X),\fm,y)\,\Big|\,p_\fm(Y)\Big],
\end{equation*}
where $p_\fm(y)=\sum_x p_\fm\big(y|do(x)\big)\cdot p(x)$ is the effective distribution on $Y$. Thus, Shannon and mutual information are simply averages of effective information, our non-universal analog of Kolmogorov complexity.

Finally, interpreting effective information as information gain, Eq.~\eqref{e:inf_gain}, and combining with results from the previous section shows that $(l -\text{ empirical VC-entropy})$ is the information we gain about the set $\Sigma_X$ of hypotheses when told that learning algorithm $\cL_{\cF,\data}$ fit the labeled data perfectly.

%%%%%%%%%%%%%%%%%%%%%%%%%%%%%%%%%%%%%%%%%%%%%%%%%%%%%%%%%%%%%%%%%%%%%%%%
\subsubsection*{Discussion}

This note starts from the observation that all physical systems classify inputs and thereby generate information. A deterministic physical system $f:X\rightarrow Y$ implicitly categorizes its inputs by assigning them to outputs: the category assigned to output $y$ is the set of inputs in the pre-image $f^{-1}(y)\subset X$. The intuition carries through in the probabilistic case after replacing pre-images with actual repertoires. Effective information then quantifies the sharpness of categories: the sharper a category, the more informative the corresponding output. Alternatively, effective information quantifies causal dependencies: outputs with high $ei$ are extremely sensitive to changes in the input.

Effective information is a concrete, computable analog of Kolmogorov complexity. The Kolmogorov complexity of a string quantifies the ``work'' required to produce it; roughly, the length of the programs that output it. Since universal Turing machines require infinite storage space and are therefore impossible to construct, it is unclear how relevant they are to processes actually occurring in nature. Effective information substitutes a deterministic model of a physical system in place of the universal Turing machine, and quantifies the ``work'' required to produce an output as the number of Y/N decisions required to choose it. 

Both Shannon and mutual information arise as expectations of effective information after tweaking to get rid of the uniform prior. The difference between Kolmogorov complexity and Shannon information reduces to: (i) replacing a universal Turing machine with a specific system (channel) and (ii) computing the average information gain over all outputs, rather than a single one. 

When the physical process under consideration is empirical risk minimization, the effective information it generates contributes to bounds on expected risk. In particular, the work (the number of Y/N decisions) required to fit data $\data$ using functions in $\cF$ essentially is the empirical VC-entropy. Since finding the optimal classifier in $\cF$ requires computing $\cL_{\cF,\data}$ in some way or another, thereby implementing it physically, it follows that the effective information generated while fitting data has implications for the future performance of classifiers, see \cite{vapnik:98, boucheron:00, bousquet:04}.

Effective information and the expected risk over the effective distribution also provide new interpretations of VC-entropy and Rademacher complexity in terms of falsifying hypotheses, see Eq.~\eqref{e:falsify1} and \eqref{e:falsify2} -- and also \cite{corfield:09} for a comparison of falsification with VC-dimension. Viewing empirical risk minimization as a physical process that classifies hypotheses according to fit $\epsilon$ thus directly links VC-entropy and Rademacher complexity with Popper's proposal that the power of a scientific theory lies in how many hypotheses \emph{it rules out}, rather than the amount of data it explains \cite{popper:59}. 

The links with Kolmogorov complexity, learning theory, information gain and falsification shown above suggest it is worth investigating whether the effective information generated while optimizing quantities other than empirical risk (e.g. margins) has implications for future performance.

\textbf{Acknowledgements.}
I thank Samory Kpotufe and Pedro Ortega for useful discussions.

{\footnotesize
% Style BST file

      % Bibliography file (usually '*.bib' ) 
}

\end{document}